\documentclass[oldversion, printer]{aa}

\usepackage[T1]{fontenc}
\usepackage{multirow}
\usepackage{tabularx}
\usepackage{psfig}
\usepackage{booktabs}
\usepackage{graphicx}
\usepackage{latexsym}
\usepackage{mathrsfs}
\usepackage{subfigure}
\usepackage{amsmath}
\usepackage{amsfonts}
\usepackage{amssymb}
\usepackage{natbib}
\usepackage{ulem}

\advance \voffset by 1.00cm\relax
\def\msun{{\,M_{\odot}}}

\def\mdot{\dot M}
\def\dotm{\dot m}

\newcommand{\der}[2]{\ensuremath{\frac{{\rm d} #1}{{\rm d} #2}}}

\newcommand{\pder}[2]{\ensuremath{\frac{\partial #1}{\partial #2}}}

\newcommand{\be}{\begin{equation}}
\newcommand{\ee}{\end{equation}}
\newcommand{\bea}{\begin{eqnarray}}
\newcommand{\eea}{\end{eqnarray}}
\newcommand{\derln}[2]{\ensuremath{\frac{{\rm d\,ln}\, #1}{{\rm d\,ln}\, #2}}}



\begin{document}

\title{Spinning up black holes with super-critical accretion flows}

\author{Aleksander S\k{a}dowski\inst{1}
 \and
Michal Bursa\inst{2}
\and
 Marek Abramowicz\inst{1,2,3,4}
\and
        W{\l}odek Klu\'zniak\inst{1}
\and
 Jean-Pierre Lasota\inst{5,6}
\and
 Rafa{\l} Moderski\inst{1}
\and
 Mohammadtaher Safarzadeh\inst{3,7}}

   \institute{
 Nicolaus Copernicus Astronomical Center, Polish Academy
             of Sciences,
             Bartycka 18, PL-00-716 Warszawa, Poland \\
             \email{as@camk.edu.pl}, ~\email{wlodek@camk.edu.pl}, ~\email{moderski@camk.edu.pl}
         \and
             Astronomical Institute, Academy of Sciences of the
            Czech Republic,
             Bo{\v c}n{\'\i} II 1401/1a, 141-31 Praha 4,
            Czech Republic\\
             \email{bursa@astro.cas.cz}
         \and
             Department of Physics, G\"oteborg University,
             SE-412-96 G\"oteborg, Sweden    \\
             \email{marek.abramowicz@physics.gu.se}
        \and
Department of Physics,
Silesian University at Opava,
Bezru{\v c}ovo n{\' a}m{\v e}st{\'\i} 1150/13,
746 01 Opava
\and
             Institut d'Astrophysique de Paris, UMR 7095 CNRS, UPMC Univ Paris 06, 98bis Bd Arago, 75014 Paris, France\\
             \email{lasota@iap.fr}
          \and
             Jagiellonian University Observatory, ul. Orla 171,
             PL-30-244 Krak{\'o}w,
             Poland
\and
     Department of Physics and Astronomy,
Johns Hopkins University, 3400 N. Charles Street
Baltimore, MD 21218\\
\email{mts@pha.jhu.edu}
            }
\date{Received ????; accepted ???? }

\abstract{
We study the process of spinning up black holes by accretion from slim disks in a wide range of accretion rates.
We show that for super-Eddington accretion rates and low values of the viscosity parameter $\alpha$ ($\lesssim 0.01$)
the limiting value of the dimensionless spin parameter $a_*$ can reach values higher than $a_*=0.9978$ inferred by Thorne (1974) in his seminal study. For $\mdot=10\mdot_{Edd}$ and $\alpha=0.01$ spin equilibrium is reached at $a_*=0.9994$. We show that the equilibrium spin value depends strongly on the assumed value of $\alpha$. We also prove that for high accretion rates the impact of captured radiation on spin evolution is negligible.
}

\authorrunning{A. S{\k a}dowski et al.}
\titlerunning{Spinning up black holes with super-critical accretion flows}
\maketitle

\section{INTRODUCTION}

Astrophysical black holes (BHs) are very simple objects -- they can be described by just two parameters: mass and angular momentum. In isolation, BHs conserve the birth values of these parameters but often, e.g. in close binaries or in active nuclei, they are surrounded by accretion disks and their mass and angular momentum change. Accretion of matter always increases BH's irreducible mass and may change its angular momentum, usually described by the dimensionless spin parameter $a_*=a/M=J/M^2$. The sign of this change and its value depend on the (relative) sign of accreted angular momentum and the balance between accretion of matter and various processes extracting BH's rotational energy and angular momentum.

The question about the maximal possible spin of an object represented by the Kerr solution of the Einstein equation is of fundamental and practical (observational) interest. First, a spin $a_* > 1$ corresponds to a {\it naked} singularity and not to a black hole. According to the Penrose cosmic censorship conjecture, naked singularities cannot form through actual physical processes, i.e. singularities in the Universe (except for the initial one in the Big Bang) are always surrounded by event horizons \citep{wald-84}. This hypothesis has yet to be proven.

In any case, the {}``third law'' of BH thermodynamics \citep{bardeenetal-73} asserts that a BH cannot be spun-up in a finite time to the extreme spin value $a_* = 1$.  
Determining the maximum value of BH spin is also of practical interest because the efficiency of accretion through a disk depends on the BH's spin value. For example, for the {}``canonical'' value $a_*=0.998$ (see below) it is about $\eta\approx 32\%$, while for $a_*\rightarrow 1$ one has $\eta\rightarrow 42\%$. Recently, \cite{banadosetal-09} showed that the energy $E_{\rm COM}$ of the center-of-mass collision of two particles colliding arbitrary close to the BH horizon, grows to infinity ($E_{\rm COM} \rightarrow \infty$) when $a_* \rightarrow 1$\footnote{Of more fundamental interest is the fact that the proper geodesic distances $D$ between the ISCO and several other special Keplerian orbits relevant to accretion disk structure tend to infinity $D \rightarrow \infty$ when $a_* \rightarrow 1$ \citep{bardeenetal-72}.}.

A definitive study of the BH spin evolution will only be possible when reliable, non-stationary models of accretion disks and jet emission mechanisms are established. For now, one has to use simplified analytical or numerical models.

\cite{thorne-74} used the model of a radiatively efficient, geometrically thin accretion disk \citep{nt} to evaluate BH spin evolution taking into account the decelerating impact of disk-emitted photons. The maximum value so obtained, $a_*=0.9978$, has been regarded as the canonical value for the maximal BH spin. In this work we generalize Thorne's approach, using models of advective, optically thick accretion disks (``slim disks'') to calculate maximum BH spin values for a large range of accretion rates. We show that for sufficiently large accretion rates they differ from the canonical value.

We start with a short discussion of previous work devoted to the BH spin evolution. In Sect.~\ref{s.tetrad} we give formulae for a general tetrad of an observer comoving with the accreting gas along the arbitrary photosphere surface. In Sect.~\ref{s.spinevolution} we give basic equations describing the BH spin evolution. The following section describes the applied model of slim accretion disks. In Sect.~\ref{s.results} we present and discuss the terminal spin values for all the models we consider. Finally, in Sect.~\ref{s.summary} we summarize our results.

\subsection{Previous studies}

A number of authors have studied the evolution of the BH spin
resulting from disk accretion. \cite{bardeen-70} initiated this field of research
by stating the problem and solving equations describing
the BH spin evolution for
accretion from the marginally stable orbit. Neglecting the effects of radiation he proved that such a process
may spin-up the BH up to $a_*=1$. Once the classical models of accretion disks were formulated
\citep{shakura-73,nt}, it was possible to account properly
for the decelerating impact of radiation (frame dragging makes counter-rotating photons more likely to be captured by the BH). As mentioned above, \cite{thorne-74} performed this study
and obtained the terminal spin value for an isotropically emitting thin disk of
$a_*=0.9978$, independently of the accretion rate.
The original study by Thorne was followed by many papers, some of which are briefly mentioned below.

The first to challenge the universality of Thorne's limit were \cite{abramowiczlasota-80} who showed that geometrically thick accretion disks may spin up BHs to terminal spin values much closer to unity than the presumably canonical $a_*= 0.9978$. Their simple argument was based on models by \cite{koz-78} who showed that for high accretion rates the inner edge of a disk may be located inside the marginally stable orbit; with increasing accretion rate, arbitrarily close to the marginally bound orbit. However, this conclusion assumed implicitly a low viscosity parameter $\alpha$, whereas for high viscosities the situation is more complicated \citep[see][and references therein]{abramowicz-inneredge}.

\cite{moderski-98} assessed the impact
of possible interaction between the disk magnetic field and the BH through
the Blandford-Znajek process. They showed that the terminal spin value may be decreased
to any, arbitrarily small value, if only the disk magnetic field is strong enough.
Given the current lack of knowledge about the strength of magnetic fields \citep[or the magnetic transport of angular momentum in the disk, see][]{ghoshabramowicz-97}
and processes leading to jet emission, a more detailed study cannot be performed. The situation may further be complicated by energy extraction from the inner parts of accretion disks \citep{livioetal-99}.

\cite{pophamgammie-98} studied the spinning-up
of BHs by optically thin advection dominated accretion flows (ADAFs). They
neglected the contribution of radiation to BH spin as such disks are radiatively
inefficient. They found that the terminal value of BH spin is very sensitive to
the assumed value of the viscosity parameter $\alpha$ and may vary between
$0.8$ and $1.0$. \cite{gammie-04}, besides making a comprehensive summary of different
ways of spinning up supermassive BHs, presented results based on a single run of a GRMHD
simulation (with no radiation included) obtaining terminal spin $a_*=0.93$.

Cosmological evolution of spins of supermassive BHs due to hierarchical mergers and thin-disk accretion episodes has been recently intensively studied. Although \cite{volonterietal-05} arrived to the conclusion that accretion tend  to spin-up BHs close to unity, as opposed to mergers which, on the average, do not influence the spin evolution, the following studies by e.g.
\citet{volonterietal-07,kingetal-08,bertivolonteri-08} showed that the situation is more complex, the final spin values depending on the details of the history of the accretion events \citep[see also][]{fanidakisetal-11}.

\cite{lietal-05} included the returning radiation into the thin-disk model of Novikov \& Thorne and calculated the spin-up limit for the BH assuming the radiation crossing the equatorial plane inside the marginally stable orbit to be advected onto the BH. Their result ($a_*=0.9983$) slightly differs from Thorne's result, thus showing that returning radiation has only a slight impact on the process of spinning-up BHs. In our study we use advective, optically thick solutions of accretion disks and account for photons captured by the BH in detail. However, we neglect the impact of the radiation returning to the disk on its structure.

\section{THE TETRAD}
\label{s.tetrad}
We base this work on slim accretion disks, which are not razor-thin and have angular momentum profile that is not Keplerian (for details on the assumptions made and the disk appearance see Sect.~\ref{s.slimdisks}). Therefore, photons are not emitted from matter in Keplerian orbits in the equatorial plane and the classical expressions for photon momenta \citep[e.g.,][]{mtw} cannot be applied. Instead, to properly describe the momentum components of emitted photons, we need a tetrad for the comoving observer instantaneously located at the disk photosphere. Below we give the explicit expression for the components of such a tetrad assuming time and axis symmetries. A detailed derivation is given in Appendix~\ref{ap.tetrad}.

Let us choose the following comoving tetrad,
\be
\label{tetrad}
e^i_{(A)}=[u^i,N_*^i,\kappa_0^i,S^i],
\ee
where
\noindent $ u^i $ is the four-velocity of matter (living in $[ t, \phi, r, \theta]$),

\vskip 0.1truecm

\noindent $ N_*^i$ is a unit vector orthogonal to the photosphere ($[r, \theta]$),

\vskip 0.1truecm

\noindent $ \kappa_0^i$ is a unit vector orthogonal to $u^i$ ($[t, \phi]$),

\vskip 0.1truecm

\noindent $ S^i$ is a unit vector orthogonal to $u^i$, $N^i$ and
$\kappa_0^i$ ($[t, \phi, r, \theta]$).

The tetrad components are given by,
\be N_*^r=\der{\theta_*}r(-g_{\theta\theta})^{-1/2}\left[1+\frac{g_{rr}}{g_{\theta\theta}}\left(\der{\theta_*}r\right)^2\right]^{-1/2},\ee
\be N_*^\theta=(-g_{\theta\theta})^{-1/2}\left[1+\frac{g_{rr}}{g_{\theta\theta}}\left(\der{\theta_*}r\right)^2\right]^{-1/2},\ee

\vspace{.5cm}

\be u^i=\frac{\eta^i+\Omega\xi^i+vS^i_*}{\sqrt{g_{tt}+\Omega g_{\phi\phi}(\Omega-2\omega)-v^2}},\ee

\vspace{.5cm}
\be \kappa_0^i=\frac{(l\eta^i+\xi^i)}{\left[-g_{\phi\phi}(1-\Omega l)(1-\omega l)\right]^{1/2}},\ee

\vspace{.5cm}
\be S^i=(1+\tilde A^2v^2)^{-1/2}(\tilde A vu^i+S^i_*),\ee
where $\theta=\theta_*(r)$ defines the location of the photosphere, $\eta_i$ and $\xi_i$ are the Killing vectors, $l=u_\phi/u_t$, $\Omega=u^\phi/u^t$, $\omega$ is the frequency of frame-dragging and the expressions for $v$ and $S^i_*$ are given in Eqs.~(\ref{e.V}) and (\ref{e.Sstar}), respectively.

\section{SPIN EVOLUTION}
\label{s.spinevolution}
\subsection{Basic equations}
The equations describing the evolution of BH dimensionless spin
parameter $a_*$ with respect to the BH energy $M$ and the accreted
rest-mass $M_0$ are \citep{thorne-74},
\be
\label{eq.spinevolution1}
\der{a_*}{{\,\rm ln} M}=\der{J/M^2}{\,{\rm ln} M}=\frac 1M\frac{\dot M_0
  u_{\phi}+\left(\der Jt\right)_{\rm rad}}{\dot M_0 u_{t}+\left(\der
    Mt\right)_{\rm rad}}-2a_*,
\ee
\be
\label{eq.spinevolution2}
\der M{M_0}=u_{t}+\left(\der Mt\right)_{\rm rad}/\dot M_0.
\ee
The energy and angular
momentum of BH increases due to the capture of photons according to
the following formulae,
\be
\label{e.dM}
(dM)_{\rm rad}=\int^{}_{disk} T^{ik}\eta_k N_i{\rm d}S,
\ee
\be
\label{e.dJ}
(dJ)_{\rm rad}=\int^{}_{disk} T^{ik}\xi_k N_i{\rm d}S,
\ee
where $\eta_k$ and $\xi_k$ are the Killing vectors connected with time
and axial symmetries, respectively, $T^{ik}$ is the stress-energy
tensor of photons, which is non-zero only for photons crossing the BH
horizon and ${\rm d}S$, the ``volume element'' in the hypersurface
orthogonal to $N^i$, is given by Eq.~\ref{ap.dS}.

From Eqs.~(\ref{e.dM}) and (\ref{e.dJ}) it follows that
\be
\left(\der  Mt\right)_{\rm rad}=\int^{2\pi}_0\int_{r_{\rm in}}^{r_{\rm out}}T^{ik}\eta_k
N_i {\rm d}\tilde S,
\label{e.dMdt1}
\ee
\be
\left(\der
  Jt\right)_{\rm rad}=\int^{2\pi}_0\int_{r_{\rm in}}^{r_{\rm out}}T^{ik}\xi_k N_i {\rm d}\tilde S,
\label{e.dJdt1}
\ee
where
\be
 {\rm d\tilde S} = {\rm d}\phi\,{\rm d}r\,\left( g_{t\phi}^2 -
g_{tt}\,g_{\phi\phi}\right )^{1/2}\,\sqrt{ g_{rr} + g_{\theta
\theta}\left(\frac{d\theta_*}{dr}\right)^2}.
\ee
\subsection{Stress energy tensor in the comoving frame}

Let us choose the tetrad given in Eq.~(\ref{tetrad}):
\begin{eqnarray}
\label{e.comtetrad}
e^i_{(0)}= u^i&&e^i_{(1)}=N^i\\\nonumber
e^i_{(2)}=\kappa^i&&e^i_{(3)}= S^i
\end{eqnarray}
The disk properties,
e.g., the emitted flux, are usually given in the comoving frame defined
by Eq.~(\ref{e.comtetrad}). The stress tensor components in the two
frames (Boyer-Lindquist and comoving) are related
in the following way,

\be
T^{ik}=T^{(\alpha)(\beta)}e^{i}_{(\alpha)}e^{k}_{(\beta)}.
\ee
The stress tensor in the comoving frame is
\be
T^{(\alpha)(\beta)}=2\int_0^{\pi/2}\int_0^{2\pi}I_0SC\pi^{(\alpha)}\pi^{(\beta)}\sin\tilde
a\,d\tilde a\,d\tilde b,
\ee
where $I_0S=I_0(r)S(\tilde a,\tilde b)$ is the intensity of the emitted
radiation, $\tilde a$ and $\tilde b$ are the angles between the emission vector and the
$N^i$ and $S^i$ vectors, respectively, $C$ is the capture function
defined in Sect.~\ref{s.capture}, the factor $2$ comes from the fact that the disk emission comes from both sides of the disk and $\pi^{(\alpha)}=p^{(\alpha)}/p^{(0)}$ are the
  normalized components of the photon four-momentum in the comoving
  frame. The latter are given by the following simple relations \citep{thorne-74},
\begin{eqnarray}\nonumber
\pi^{(0)}&=&1,\\
\pi^{(1)}&=&\cos \tilde a,\\\nonumber
\pi^{(2)}&=&\sin{\tilde a}\cos \tilde b,\\\nonumber
\pi^{(3)}&=&\sin{\tilde a}\sin \tilde b.
\end{eqnarray}
Eqs.~(\ref{e.dMdt1}) and (\ref{e.dJdt1}) take the form,
\be
\left(\der Mt\right)_{\rm rad}=\int^{}_{\rm disk}
T^{(\alpha)(\beta)}e^{i}_{(\alpha)}e^{k}_{(\beta)}\eta_k N_i{\rm
  d}\tilde S,
\label{e.dMdt2}
\ee
\be
\left(\der Jt\right)_{\rm rad}=\int^{}_{\rm disk}
T^{(\alpha)(\beta)}e^{i}_{(\alpha)}e^{k}_{(\beta)}\xi_k N_i{\rm
  d}\tilde S,
\label{e.dJdt2}
\ee
where $e^i_{(a)}$ is our local frame tetrad given by
Eq.~(\ref{e.comtetrad}). Taking the following relations into account,
\bea
\pi^{(\alpha)}&=&\pi^je^{(\alpha)}_j,\\\nonumber
e^{(\alpha)}_je_{(\alpha)}^i&=&\delta_j^i,
\eea
we have,
\vspace{.3cm}
\be \pi^{(\alpha)}e^{i}_{(\alpha)}N_i=\pi^{(\alpha)}\delta^{(1)}_{(\alpha)}=\pi^{(1)}=\cos \tilde a,\ee
\be \pi^{(\beta)}e^{k}_{(\beta)}\eta_k=\pi^je^{(\beta)}_ju^k_{(\beta)}\eta_k=\pi^j\delta_j^k\eta_k=\pi^k\eta_k=\pi_t,\ee
\be \pi^{(\beta)}e^{k}_{(\beta)}\xi_k=\pi^je^{(\beta)}_ju^k_{(\beta)}\xi_k=\pi^j\delta_j^k\xi_k=\pi^k\xi_k=\pi_\phi,\ee
where,
\bea
\pi_t&=&\pi^{(i)}e^t_{(i)} g_{tt}+\pi^{(i)}e^\phi_{(i)}
g_{t\phi},\\
\pi_\phi&=&\pi^{(i)}e^t_{(i)} g_{t\phi}+\pi^{(i)}e^\phi_{(i)} g_{\phi\phi}.
\eea
Therefore, Eqs.~(\ref{e.dMdt2}) and (\ref{e.dJdt2}) may be finally expressed as,
\begin{eqnarray}\nonumber
\left(\der
  Mt\right)_{\rm rad}&=&4\pi\int_{r_{\rm in}}^{r_{\rm out}}\int_0^{2\pi}\int_0^{\pi/2}I_0SC\pi_t\times
\\
&\times&\cos\tilde
a\sin\tilde a\,d\tilde a\,d\tilde b \sqrt{\tilde g}\,{\rm d}r,
\end{eqnarray}
\begin{eqnarray}\nonumber
\left(\der
Jt\right)_{\rm rad}&=&4\pi\int_{r_{\rm in}}^{r_{\rm out}}\int_0^{2\pi}\int_0^{\pi/2}I_0SC\pi_\phi\times
\\
&\times&\cos\tilde
a\sin\tilde a\,d\tilde a\,d\tilde b \sqrt{\tilde g}\,{\rm d}r,
\end{eqnarray}
with
\be \sqrt{\tilde g}\equiv\left( g_{t\phi}^2 -
g_{tt}\,g_{\phi\phi}\right )^{1/2}\,\sqrt{ g_{rr} + g_{\theta
\theta}\left(\frac{d\theta_*}{dr}\right)^2}.\ee

\subsection{Emission}
The intensity of local radiation may be identified with the flux
emerging from the disk surface,
\be I_0=F(r).\ee
The angular emission factor $S$ is given by \citep{thorne-74},
\begin{equation}
S(\tilde a,\tilde b)=\left\{\begin{array}{ll}
1/\pi & {\rm isotropic}\\
(3/7\pi)(1+2\cos\tilde a)& {\rm limb\,darkening}\\
\end{array}\right.
\end{equation}
for isotropic and limb-darkened cases, respectively. In this work we assume that the
radiation is emitted isotropically.
\subsection{Capture function}
\label{s.capture}
The BH energy and angular momentum are affected only by photons
crossing the BH horizon. Following \cite{thorne-74}, we define
the capture function $C$,
\begin{equation}
C=\left\{\begin{array}{ll}
1& {\rm if\ the\ photon\ hits\ the\ BH,}\\
0& {\rm in\ the\ opposite\ case.}\\
\end{array}\right.
\end{equation}
Herein, we calculate $C$ in two ways. First, we use the original
\cite{thorne-74} algorithm modified to account for emission out
of the equatorial plane. For this purpose we calculate
the constants of motion, $j$ and $k$, for a geodesic orbit of a photon in
the following way,
\be j=a_*^2+a_*(\pi_\phi/M\pi_t),\ee
\be k=\frac1{(M\pi_t)^2}\left[\pi_\theta^2-(\pi_\phi+a_*M\pi_t \sin\theta_*)^2/\sin^2\theta_*\right]\ee
which replaces Thorne's Eqs.~(A10). This approach does not take into
account possible returning radiation, i.e., a photon hitting the disk
surface is assumed to continue its motion. Such a treatment is not appropriate
for optically thick disks - returning photons are most likely
absorbed or advected towards the BH.

To assess the importance of this inconsistency we adopt two additional
algorithms for calculating $C$. Using photon equations of motion we determine if the photon hits the disk surface \citep{bursa.raytracing}. Then we make two assumptions, either the angular momentum and energy of all ``returning`` photons are advected onto the BH ($C_1$), or all are re-emitted carrying away their original angular momentum and energy, and never hit the BH ($C_2$). In this way we establish two limiting cases allowing us to assess the impact of the returning radiation.

We note here that for fully consistent treatment of the returning
radiation \citep[as in][for geometrically thin disks]{lietal-05} it is not enough to modify the capture function,
solving for the whole structure of a self-irradiated accretion disk
is necessary. The latter has not yet been done for luminous and
geometrically thick disks. We are currently working on implementing such
a scheme and will study its impact onto the BH spin evolution in a
forthcoming paper.

\section{SLIM ACCRETION DISKS}
\label{s.slimdisks}
\subsection{Equations}
In this section we present slim disk equations.
They were derived originally by \cite{lasota-94} and improved e.g., by
\cite{adafs} and \cite{gammie}. Here, we follow \cite{katobook} and
assume the polytropic equation of state when performing vertical
integration. The formalism we use in this section was adopted from
\cite{sadowski-vertslim}.

In the structure equations we take $G=c=1$, and make use of the following expressions involving the BH spin:
\begin{eqnarray}\nonumber
\Delta&=&r^2-2Mr+a^2,\\\nonumber
A&=&r^4+r^2a^2+2Mra^2,\\
\cal C&=&1-3r_*^{-1}+2a_*r_*^{-3/2}\\\nonumber
\cal D&=&1-2r_*^{-1}+2a_*^2r_*^{-2}\\\nonumber
\cal H&=&1-4a_*r_*^{-3/2}+3a_*^{2}r_*^{-2} \nonumber
\end{eqnarray}
with $a_*=a/M$ and $r_*=r/M$. 

We also define
\be
\Omega_\perp^2\equiv\frac{M}{r^3}\frac{\cal H}{\cal C}
\ee
and dimensionless accretion rate
${\dot m}= {\dot M}/{\dot M}_{\rm Edd}$, where ${\dot M}_{\rm Edd}
= 16L_{\rm Edd}/c^2$ is the critical accretion rate approximately
corresponding to the Eddington luminosity ($L_{\rm Edd} =
1.25\times10^{38}M/M_{\odot}\,$erg/s) for a disk around a
non-rotating black hole.

The equations describing slim disks,
written in the cylindrical coordinates are:

(i) Mass conservation:
\begin{equation}
  \dot M=-2\pi \Sigma \Delta^{1/2}\frac{v}{\sqrt{1-v^2}},
\label{eq_cont2}
\end{equation}
where $\Sigma=\int_{-h}^{+h}\rho \,dz$ is disk surface density, while $v$ denotes gas velocity as measured by an observer co-rotating
with the fluid and is related to the four-velocity $u^r$ by $R u^r=\Delta^{1/2}v/\sqrt{1-v^2}$

(ii) Radial momentum conservation:
\begin{equation}
\frac{v}{1-v^2}\frac{dv}{dR}=\frac{\cal A}{R}-\frac{1}{\Sigma}\frac{dP}{dR},
\label{eq_rad3}
\end{equation}
where
\begin{equation}
{\cal A}=-\frac{MA}{R^3\Delta\Omega_k^+\Omega_k^-}\frac{(\Omega-\Omega_k^+)(\Omega-\Omega_k^-)}{1-\tilde\Omega^2\tilde R^2},
\label{eq_rad4}
\end{equation}
and $\Omega=u^\phi /u^t$ is the angular velocity with respect to a stationary observer, $\tilde\Omega=\Omega-\omega$ is the angular velocity with respect to an inertial observer, $\Omega_k^\pm=\pm M^{1/2}/(R^{3/2}\pm aM^{1/2})$ are the angular frequencies of the co-rotating and counter-rotating Keplerian orbits and $\tilde R=A/(R^2\Delta^{1/2})$ is the radius of gyration.
$P=\int_{-h}^{+h}p\,dz$ is the vertically integrated total pressure.

(iii) Angular momentum conservation:
\begin{equation}
 \frac{\dot{M}}{2\pi}({\cal L}-{\cal L}_{\rm in})=\frac{A^{1/2}\Delta^{1/2}\Gamma}{R}\alpha P,
\label{eq_ang6}
\end{equation}
where ${\cal L}=u_\phi$, ${\cal L}_{\rm in}$ is a constant and
$\Gamma$ is the Lorentz factor \citep{gammie}:
\be \Gamma^2=\frac1{1-v^2}+\frac{{\cal L}^2r^2}{A}.\ee

(iv) Vertical equilibrium:
\be
H^2\Omega_\perp^2=(2N+3)\frac P\Sigma.
\ee

(v) Energy conservation:

\begin{eqnarray}\nonumber
\label{eq.qadv}
Q^{\rm adv}&=&-\frac{\mdot}{2\pi R^2}\left(\eta_3\frac{P}\Sigma\derln{P}{R} - (1+\eta_3)\frac{P}\Sigma\derln\Sigma R+\right.\\
&+&\left. \eta_3\frac P\Sigma\derln{\eta_3}{R}+\Omega^2_\perp\eta_4\derln{\eta_4}{R}\right),
\end{eqnarray}
where the amount of heat advected $Q^{\rm adv}$ is:
\be
Q^{\rm adv}=-\alpha P\frac{A\gamma^2}{R^3}\der\Omega r - f_F\frac{64\sigma T_C^4}{3\Sigma\kappa}.
\ee
Assuming the polytropic index $N=3$ we have:
\begin{eqnarray}
\eta_1&=&\frac1{T_0^4}\int_0^HT^4\,dz=128/315\, H\\\nonumber
\eta_2&=&\frac2{\Sigma T_0}\int_0^H\rho T\,dz=40/45\\\nonumber
\eta_3&=&\frac1P\left(\frac1{5/3-1}\frac{k}\mu\frac{40}{45}\Sigma T_C+\frac{256}{315}aT^4_CH\right)\\\nonumber
\eta_4&=&\frac1{\Sigma}\int_0^H\rho z^2\,dz=1/18\, H^2.
\end{eqnarray}

\noindent The equations given above form a two-dimensional system of ordinary differential
equations with a critical (i.e., sonic) point. For each
set of disk parameters the regular solution is possible only for one specific
value of ${\cal L}_{\rm in}$, which is an eigenvalue of the problem. The appropriate value
may be found using either the relaxation or the shooting method. For details on the numerical procedures
see \cite{sadowski-slim} and \cite{sadowski-vertslim}.

\subsection{Disk appearance}

In this section we briefly describe the properties of slim disk
solutions. For a more detailed discussion see e.g., \cite{sadowski-slim},
\cite{abramowicz-inneredge}, or \cite{bursa-slimbb}.

\begin{figure}
\centering
  \includegraphics[width=.48\textwidth,angle=0]{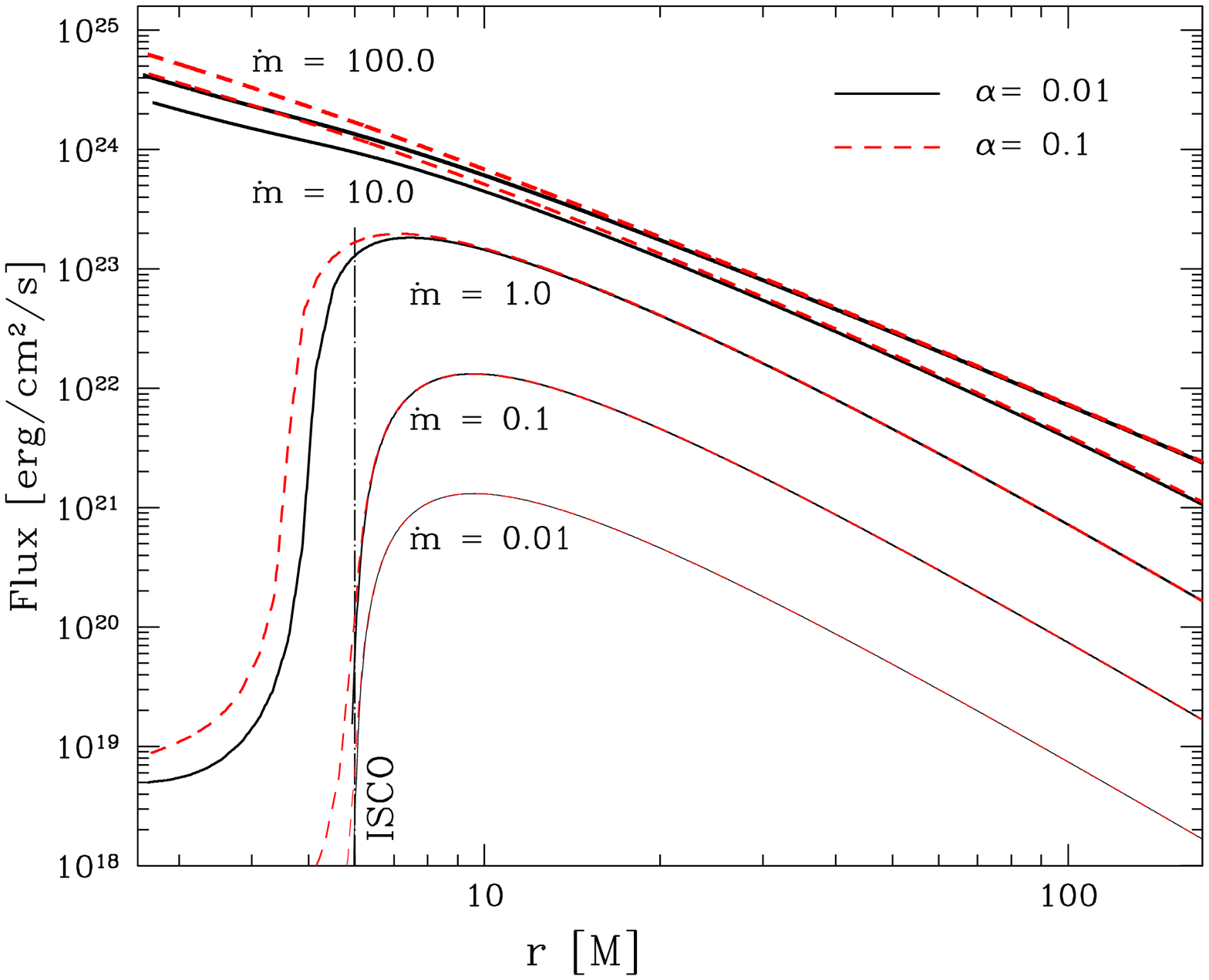}

  \caption{Flux profiles for $M=10\msun$ and $a_*=0.0$.}
  \label{f.flux}
\end{figure}

The radial profiles of the emitted flux for a non-rotating BH
 are presented in Fig.~\ref{f.flux}. For low accretion rates
($\dot m\ll 1$) they almost coincide with the Novikov \& Thorne
solutions (the small departure is due to the angular momentum taken
away by photons, an effect which is neglected in our slim disk scheme).
When the accretion rate becomes high advective
cooling starts to play a significant role and the emission departs
from that of the radiatively efficient solution. This departure is visible as early as
for $\dot m=1$ --- the emission extends significantly inside the marginally
stable orbit. For super-critical accretion rates the flux increases monotonically towards the BH horizon.
Different colors in Fig.~\ref{f.flux} denote solutions
for different values of the viscosity parameter $\alpha$. Although they
are very similar, one can notice that the higher the value of $\alpha$,
the lower the accretion rate at which advection starts to modify the
emission profile.

\begin{figure}
\centering
  \includegraphics[width=.48\textwidth,angle=0]{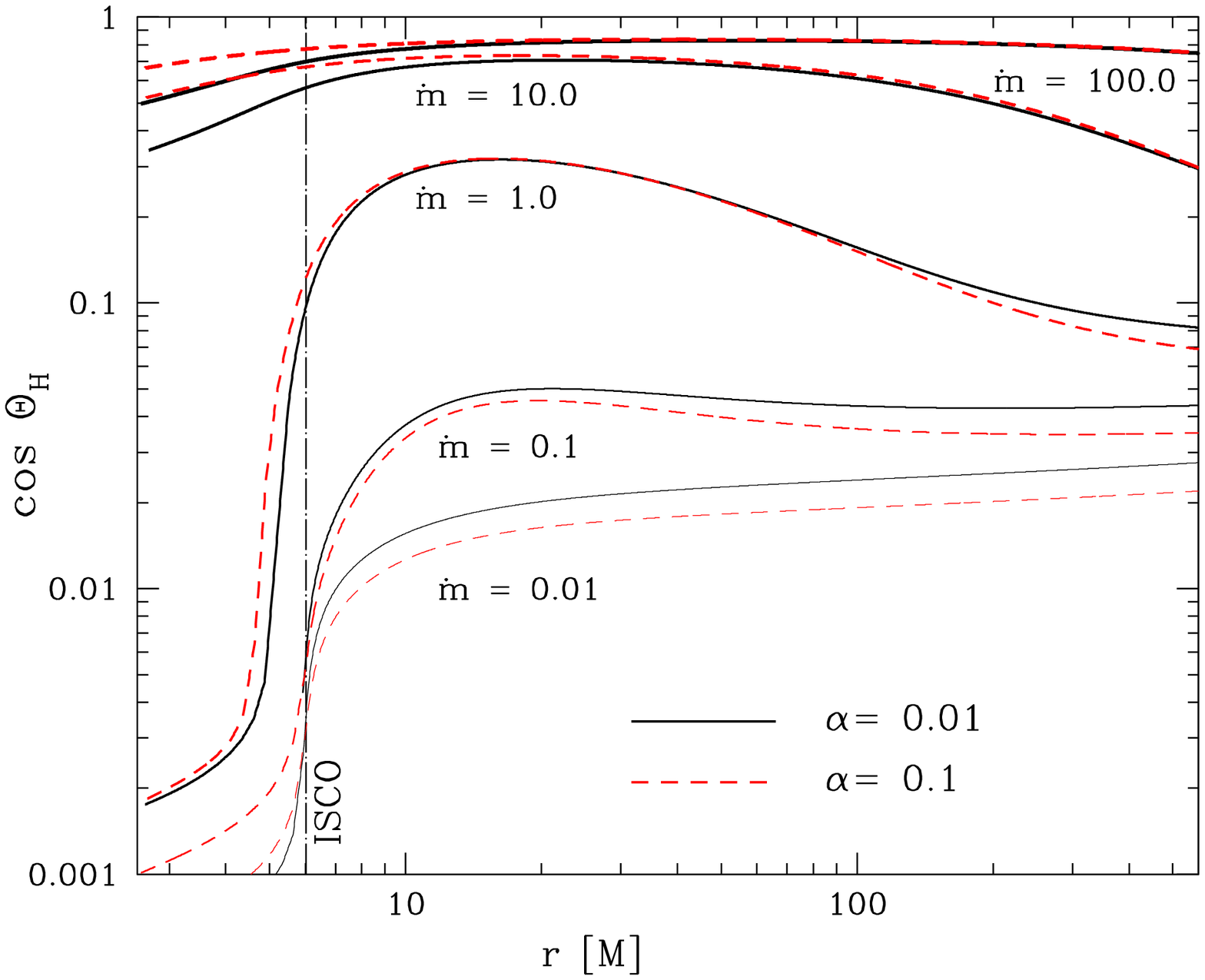}

  \caption{Photosphere profiles profiles for $M=10\msun$ and $a_*=0.0$.}
  \label{f.phot}
\end{figure}

In Fig.~\ref{f.phot} we plot disk thickness profiles ($\cos\Theta_H = H /r$) for
a range of accretion rates and two values of $\alpha$. For $\dot m>0.1$
the inner region of the accretion disk is puffed up by the radiation
pressure and the disk surface corresponds to the location
where the radiation pressure force (proportional to the local flux
of emitted radiation) is balanced by the vertical component of
the gravity force. For the Eddington luminosity ($\dot m\approx 1$)
the highest $H/R$ ratio equals $\sim 0.3$ ($\cos\Theta_H\approx0.3$),
 while for the largest accretion
rate considered ($\dot m=100$) it reaches $\sim 1.5$ ($\cos\Theta_H\approx0.83$).

\begin{figure*}
\centering
 \subfigure
{
\includegraphics[height=.4\textwidth]{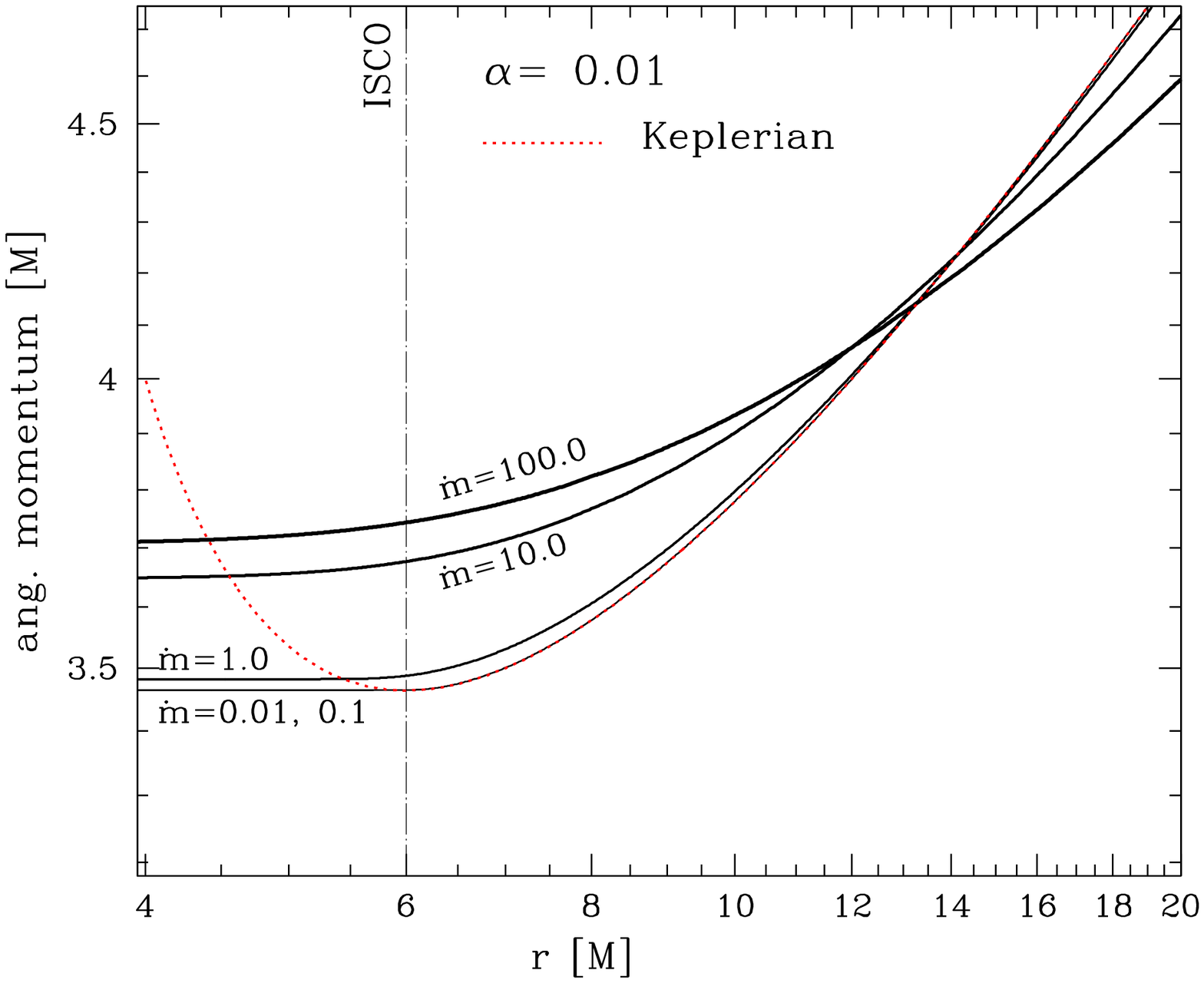}
}
\hspace{-.07\textwidth}
 \subfigure
{
\includegraphics[height=.4\textwidth]{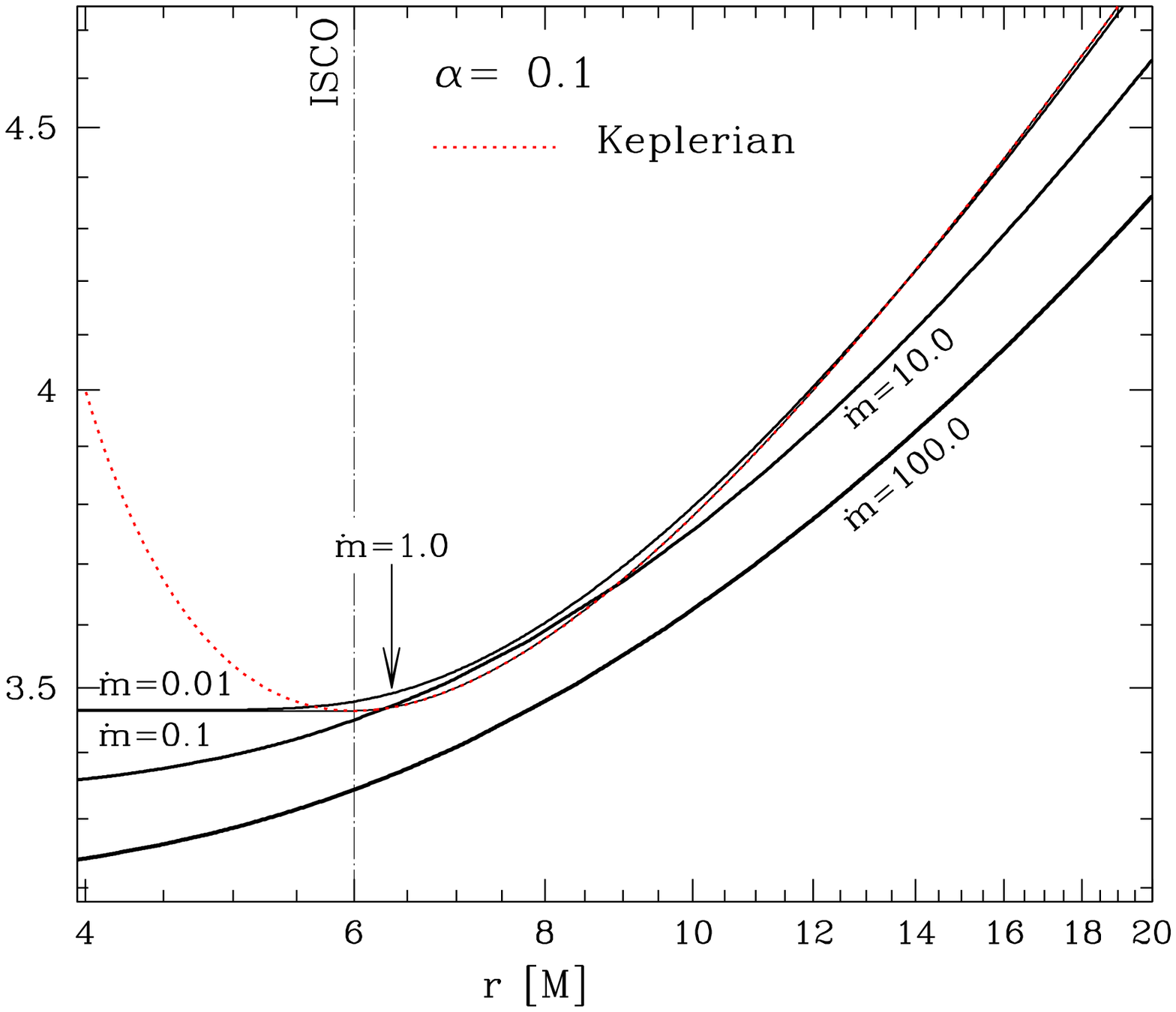}
}
\caption{ Profiles of the disk angular momentum for $\alpha=0.01$ (left)
and $\alpha=0.1$ (right panel) for different accretion rates. The spin
of the BH $a_*=0$.}
\label{f.angmom}
\end{figure*}

In the thin disk approximation the angular momentum of gas follows
the Keplerian profile. This condition is not satisfied for advective
accretion disks with significant radial pressure gradients. In Fig.~\ref{f.angmom}
we present angular momentum profiles for disks with different accretion rates,
$\alpha=0.01$ (left) and $\alpha=0.1$ (right panel). It is clear
that the higher the accretion rate, the larger the departure
from the Keplerian profile. However, the quantitative behavior
depends strongly on $\alpha$. For $\alpha\lesssim0.01$ the flow
is super-Keplerian in the inner part (e.g., between $r=4.5M$ and $r=14M$
for $\dotm=100$). For larger viscosities ($\alpha\gtrsim0.1$) and
high accretion rates the flow is sub-Keplerian at all radii.
As a result, the value of the angular momentum at the BH horizon (${\cal L}_{\rm in}$)
also depends strongly on $\alpha$, decreasing with increasing $\alpha$. Dependence of the flow
topology on the viscosity parameter has been recently studied in
detail by \cite{abramowicz-inneredge}.


\section{RESULTS FOR BH SPIN EVOLUTION}
\label{s.results}
Using the slim disk solutions described in the previous section we solve Eqs.~(\ref{eq.spinevolution1}) and (\ref{eq.spinevolution2}) using regular Runge-Kutta method of the 4th order. To calculate the integrals (Eqs.~(\ref{e.dMdt1}) and (\ref{e.dJdt1})) we use the alternative extended Simpson's rule \citep{numericalrecipes} basing on 100 grid points in $\tilde a$, $\tilde b$ and radius $r$. We have made convergence tests proving this number is sufficient.

\begin{figure}
\centering
  \includegraphics[width=.48\textwidth,angle=0]{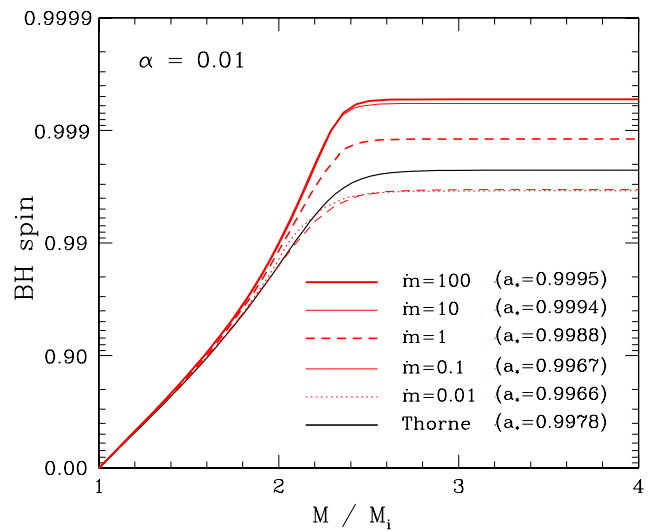}

  \caption{Spin evolution for $\alpha=0.01$.}
  \label{f.sev0.01}
\end{figure}

\begin{figure}
\centering
  \includegraphics[width=.48\textwidth,angle=0]{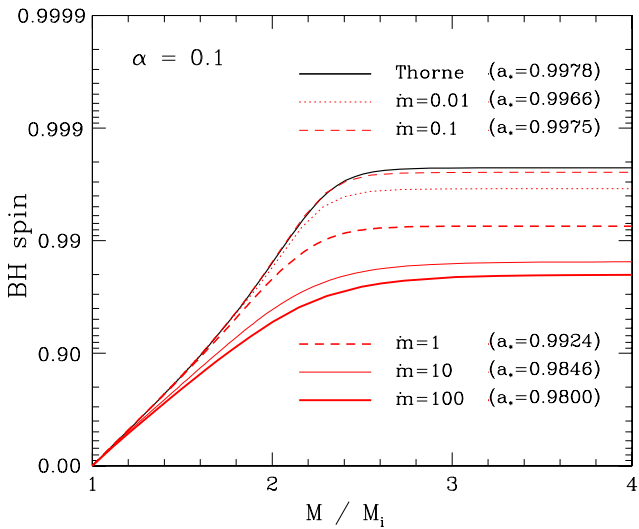}
  \caption{Spin evolution for $\alpha=0.1$.}
  \label{f.sev0.1}
\end{figure}

In Figs.~\ref{f.sev0.01} and \ref{f.sev0.1} we present the BH spin evolution for $\alpha=0.01$ and $0.1$, respectively. The red lines show the results for different accretion rates while the black line shows the classical \cite{thorne-74} solution based on the \cite{nt} model of thin accretion disk. Our low accretion rate limit does not perfectly agree with the black line as the slim disk model does not account for the angular momentum carried away by radiation. As a result, the low-luminosity slim disk solutions slightly overestimate (by no more than few percent) the emitted flux leading to stronger deceleration of the BH by radiation --- the \cite{thorne-74} result is the proper limit for the lowest accretion rates. When the accretion rate is high enough (e.g., $\dot m>0.1$), the impact of the omitted angular momentum flux is overwhelmed by the modification of the disk structure introduced by advection.

It is clear the spin evolution is quantitatively different for different values of the viscosity parameter. For the lower value ($\alpha = 0.01$) the BH spin can reach values significantly higher than $0.998$ for the highest accretion rates ($a_*=0.9995$ for $\dotm=100$), while for the higher viscosity ($\alpha = 0.1$) the terminal spin value decreases with increasing accretion rate down to $a_*=0.9800$ for $\dotm=100$. This behavior is connected with the viscosity impact on the critical point topology. Generally speaking, the higher the value of $\alpha$, the lower the angular momentum of the flow at BH horizon for a given accretion rate (Fig.~\ref{f.angmom}), leading to a slower acceleration of the BH rotation.

\begin{figure}
\centering
  \includegraphics[width=.48\textwidth,angle=0]{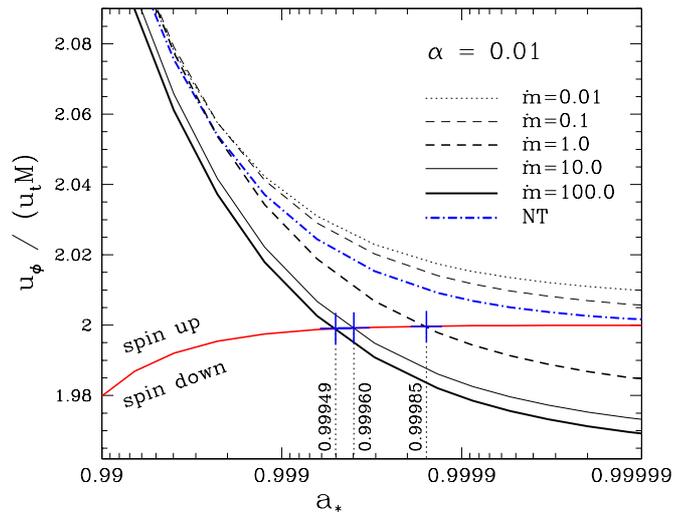}
  \caption{The rate of spin-up or spin-down by ''pure'' accretion
    (radiation neglected) for $\alpha=0.01$. Profiles for five
    accretion rates are presented. Their intersections
 with the red line (marked with blue
    crosses) correspond to equilibrium states. For the two lowest
    accretion rates the equilibrium state is never reached
    ($a_*\rightarrow 1$).}
  \label{f.eqalp0.01}
\end{figure}

\begin{figure}
\centering
  \includegraphics[width=.48\textwidth,angle=0]{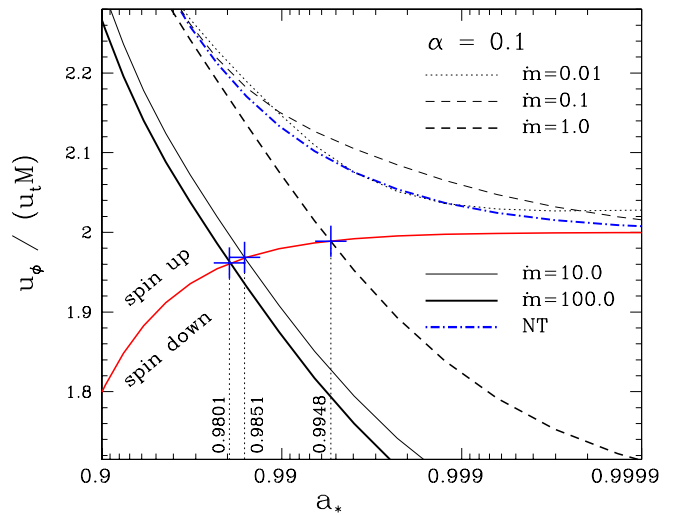}
  \caption{Same as Fig.~\ref{f.eqalp0.01} but for $\alpha=0.1$.}
  \label{f.eqalp0.1}
\end{figure}

\begin{figure}
\centering
  \includegraphics[width=.48\textwidth,angle=0]{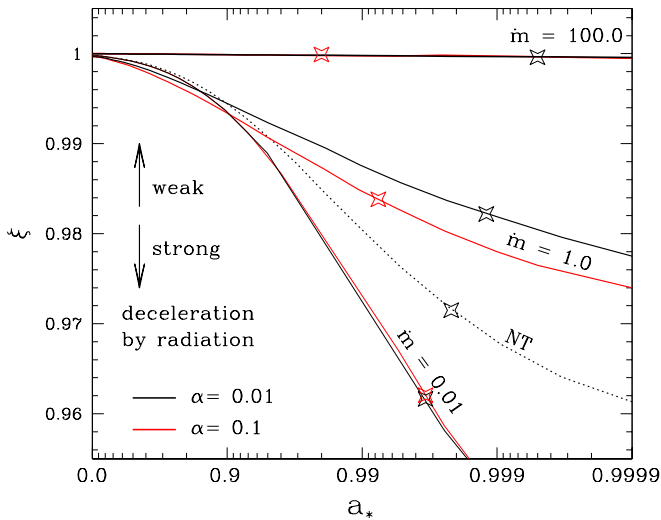}
  \caption{Radiation impact factor $\xi$ (Eq.~[\ref{e.xi}]) for different accretion rates and values of $\alpha$. The dotted line corresponds to the thin-disk induced spin evolution. For $\xi\approx1$ spin evolution is not affected by radiation. Stars denote the equilibrium states (compare Table~\ref{t.astars}).}
  \label{f.radimp}
\end{figure}

To study this fact in detail we calculated the rate of BH spin-up for ''pure'' accretion of matter (without accounting for the impact of radiation). For that case the BH spin evolution is given by (compare Eq.~(\ref{eq.spinevolution1})):
\begin{equation}
\label{eq.spinevpure}
\der{a_*}{{\rm ln} M}=\frac 1M\frac{
  u_{\phi}}{ u_{t}}-2a_*
\end{equation}
In Figs.~\ref{f.eqalp0.01} and \ref{f.eqalp0.1} we plot with black lines the first term on the right hand side of the above equation for different accretion rates and values of $\alpha$. The red lines on these plots show the absolute value of the second term. The intersections of the black and red lines denote the equilibrium states i.e., the limiting values of BH spin for pure accretion. These values differ significantly from the previously discussed results only for low accretion rates. In contrast, for high accretion rates radiation has little impact on the spin evolution and the value of terminal spin is mostly determined by the properties of the flow. In Fig.~\ref{f.radimp} we plot the radiation impact parameter $\xi$, defined as the ratio of the disk-driven terms on the right hand sides of Eqs.~(\ref{eq.spinevolution1}) and (\ref{eq.spinevpure}),
\be
\label{e.xi}
  \xi = \left.\frac{\dot M_0
  u_{\phi}+\left(\der Jt\right)_{\rm rad}}{\dot M_0 u_{t}+\left(\der
    Mt\right)_{\rm rad}}\right/ \frac{
  u_{\phi}}{ u_{t}}
\ee
 If the captured radiation significantly decelerates the BH spin-up, this ratio drops below unity. On the other hand, it is close to unity for BH spin evolution which is not affected by the radiation. According to Fig.~\ref{f.radimp} the latter is in fact the case for the highest accretion rates independently of $\alpha$.

In Table~\ref{t.astars} we list the resulting values of the terminal BH spin for all the models considered. The first column gives results for our fiducial model ($A$) including Thorne's capture function, emission from the photosphere at the appropriate radial velocity.

The second column presents results obtained assuming the same (Thorne's) capture function and profiles of emission, angular momentum and radial velocity as in the model A, but assuming the emission takes place from the equatorial plane instead of the photosphere. The resulting terminal spin values are equal, up to 4 decimal digits, to the values obtained with the fiducial model. This result is as expected for the lowest accretion rates, where the photosphere is located very close to the equatorial plane. For the highest accretion rates the location of the emission has no impact on the BH spin-up, as the spin evolution is driven by the flow itself and the effects of radiation are negligible. However, for moderate accretion rates one could have expected significant change of the terminal spin. We find that the location of the photosphere has little impact on the resulting BH spin regardless of the accretion rate.

Our third model ($V$) neglects the flow radial velocity when the radiative terms are evaluated. Similar arguments to those given in the previous paragraph apply.  For the lowest accretion rates, the radial velocity is negligible and therefore it should have no impact on the resulting spin. For the highest accretion rates, the spin-up process depends only on the properties of the flow. Once again, however, the impact of this assumption on moderate accretion rates is not obvious. The radial velocity turns out to be of little importance for the calculation of the terminal spin (only for $\alpha=0.1$ and moderate accretion rates the difference between models $A$ and $V$ is higher than $0.01\%$).

In the fourth and fifth columns of Table~\ref{t.astars}, results for models with the same assumptions as the fiducial model, but with different capture functions are presented. The first alternative capture function ($C_1$) assumes that the angular momentum and energy of all photons returning to the disk are added to that of the BH. This assumption has a strong impact on the spin evolution---the terminal spin values are higher, sometimes approaching $a_*=1$. This may seem surprising because in the classical approach the captured photons are responsible for decelerating the spin-up. This effect comes from the fact that the cross-section (with respect to the BH) of photons going ``against`` the frame dragging is larger than of photons following the BH sense of rotation. As frame dragging is involved, this effect is significant only in the vicinity of the BH horizon. For our model $C_1$, however, the probability of photons returning to the disk does not appreciably differ for co- and counter-rotating photons, as they both hit the disk surface mostly at large radii.


The other capture function ($C_2$) assumes, on the contrary, that all returning photons are re-emitted from the disk with their original angular momentum and energy (and never fall onto the BH). This assumption cuts off the photons which would hit the BH in the fiducial model after crossing the disk surface. Thus, we may expect decreased radiative deceleration and increased values of the terminal spin parameter. However, these changes are not significant, reflecting the fact that most of the photons hit the BH directly, along slightly curved trajectories. Only for $\alpha=0.1$ and moderate accretion rates do the terminal spin values differ in the $\rm 4^{th}$ decimal digit.

Neither of the models with a modified capture functions is self-consistent. To account properly for the returning radiation one has to modify the disk equations by introducing appropriate terms for the outgoing and incoming fluxes of angular momentum and additional radiative heating. No such model for advective, optically thick accretion disk has been constructed. The emission profile should be significantly affected (especially inside the marginally stable orbit) by the returning radiation leading to different rates of deceleration by photons. In view of our results for models $C_1$ and $C_2$, as well as the results of \cite{lietal-05}, one may expect the final spin values for super-critical accretion flows to be slightly higher than the ones obtained in this work.

The last column of Table~\ref{t.astars} gives terminal spin values for ''pure'' accretion (radiation neglected). Under such assumptions the BH spin could reach $a_*=1$ for sub-Eddington accretion rates as there are no photons which could decelerate and stop the spin-up process. As discussed above, for the highest accretion rates the resulting BH spin values agree with the values obtained for the fiducial model as radiation has little impact on spin evolution in this regime.

\begin{table*}
\caption{BH spin terminal values\vspace{.3cm}}
\centering
\begin{tabular}{lccccccc}
\hline \hline
\multicolumn{2}{l}{capture function:} & $C$ & $C$ & $C$ & $C_1$ & $C_2$ & - \\
\hline
\multicolumn{2}{l}{model:} & $\bf{A}$ & $T$ & $V$ & $A$ & $A$ & \textit{NR} \\
\hline \hline
\multicolumn{2}{l}{thin disk}   & \textbf{0.9978} & 0.9978 & 0.9978 & 0.9981 & 0.9978 & $\rightarrow 1$\\
\hline
\multirow{5}{*}{$\alpha=0.01$} &
  $\dot m=0.01$  & \textbf{0.9966} & 0.9966 & 0.9966 & 0.9975 & 0.9966 & $\rightarrow 1$\\
& $\dot m=0.1$   & \textbf{0.9967} & 0.9967 & 0.9967 & $\rightarrow 1$ & 0.9967 & $\rightarrow 1$\\
& $\dot m=1$     & \textbf{0.9988} & 0.9988 & 0.9988 & $\rightarrow 1$ & 0.9988 & 0.9998 \\
& $\dot m=10$    & \textbf{0.9994} & 0.9994 & 0.9994 & $\rightarrow 1$ & 0.9994 & 0.9996 \\
& $\dot m=100$   & \textbf{0.9995} & 0.9995 & 0.9995 & $\rightarrow 1$ & 0.9995 & 0.9995 \\
\hline
\multirow{5}{*}{$\alpha=0.1$} &
  $\dot m=0.01$  & \textbf{0.9966} & 0.9966 & 0.9966 & 0.9975 & 0.9966 & $\rightarrow 1$\\
& $\dot m=0.1$   & \textbf{0.9975} & 0.9975 & 0.9975 & $\rightarrow 1$ & 0.9975 & $\rightarrow 1$\\
& $\dot m=1$     & \textbf{0.9924} & 0.9924 & 0.9923 & $\rightarrow 1$ & 0.9927 & 0.9948 \\
& $\dot m=10$    & \textbf{0.9846} & 0.9846 & 0.9845 & 0.9901 & 0.9847 & 0.9951 \\
& $\dot m=100$   & \textbf{0.9800} & 0.9800 & 0.9800 & 0.9803 & 0.9800 & 0.9801 \\
\hline \hline
\label{t.astars}
\end{tabular}
\tablefoot{$\, C$ - Thorne's capture function, $C_1$ - all
returning photons advected onto the BH, $C_2$ - all returning photons neglected;
$A$ - our fiducial model, $T$ - emission from the equatorial plane, $V$ - zero radial
velocity, \textit{NR} - pure accretion, radiation neglected.}
\end{table*}

\section{SUMMARY}
\label{s.summary}
We have studied BH spin evolution due to disk accretion assuming that the angular momentum and energy carried by the flow and the emitted photons is the only process affecting the BH rotation. We generalized the original study of \cite{thorne-74} to high accretion rates by applying a relativistic, advective, optically thick slim accretion disk model. Assuming isotropic emission (no limb darkening) we have shown that

(i) the terminal value of BH spin depends on the accretion rate for $\dot m\gtrsim 1$,

(ii) the terminal spin value is very sensitive to the assumed value of the viscosity parameter $\alpha$ --- for $\alpha\lesssim 0.01$ the BH is spun up to $a_*>0.9978$ for high accretion rates, while for $\alpha\gtrsim 0.1$ to $a_*<0.9978$,

(iii) with a low value of $\alpha$ and high accretion rates, the BH may be spun up to spins significantly higher than the canonical value $a_*=0.9978$ (e.g., to $a_*=0.9994$ for $\alpha=0.01$ and $\dotm =10$) but, under reasonable assumptions, BH cannot be spun up arbitrarily close to $a_*=1$,

(iv) BH spin evolution is hardly affected by the emitted radiation for high ($\dotm\gtrsim 10$) accretion rates (the terminal spin value is determined by the flow properties only),

(v) for all accretion rates, neither the photosphere profile nor the profile of radial velocity significantly affects the spin evolution.

We point out that the inner edge of an accretion disk cannot be uniquely defined for a super-critical accretion \citep{abramowicz-inneredge}, as opposed to geometrically thin disks where the inner edge is uniquely located at the marginally stable orbit ($R_{ms}$). In the thin-disk case the BH spin evolution is determined by the flow properties at this particular radius (as there is no torque between the marginally stable orbit and BH horizon) and the profile of emission (terminating at $R_{ms}$). For super-critical accretion rates, however, one cannot distinguish a particular inner edge which is relevant to studying BH spin evolution. On the one hand, the values of the specific energy ($u_t$) and the angular momentum ($u_\phi$) remain constant within the \textit{stress inner edge}. On the other, the radiation is emitted outside the \textit{radiation inner edge}. These inner edges do not coincide as they are related to different physical processes.

Our study was based on a semi-analytical, hydrodynamical model of an accretion disk which makes a number of simplifying assumptions like stationarity, no returning radiation, $\alpha P$ prescription and no wind outflows. Thus, the results obtained in this work should be considered only qualitative, as the precise values of the terminal spin parameter are very sensitive (e.g., through viscosity) to the flow and emission properties. Moreover, we neglected the impact of magnetic fields and jet ejection mechanisms. Nevertheless, our study shows that Thorne's canonical value for BH spin ($a_*=0.9978$) may be exceeded under certain conditions.

\acknowledgements{
This work was supported in part by Polish Ministry of Science grants N203
0093/1466, N203 304035, N203 380336, N N203 381436. JPL
acknowledges support from the French Space Agency CNES, MB from ESA
PECS project No. 98040.
}

\bibliographystyle{apj}

\begin{appendix}

\section{The tetrad for an observer instantaneously located at the photosphere}
\label{ap.tetrad}

\noindent Our aim is to derive the tetrad of an observer moving along the photosphere that
would depend only on the quantities which are calculated in accretion disk models, i.e., on
the radial and azimuthal velocities of gas and the location of disk photosphere.

\noindent The metric considered here is the Kerr geometry $g_{ik}$
in the Boyer-Lindquist coordinates $[t, \phi, r, \theta]$. The
signature adopted is $+---$. Similarly as in Carter's Les Houches
lectures \citep{leshouches}, we will consider two fundamental planes; the symmetry
plane ${\cal S}_0 = [t, \phi]$ and the meridional plane ${\cal
M}_* = [r, \theta]$. (Four)-Vectors that belong to the plane ${\cal
S}_0$, will be denoted by the subscript $0$, and vectors that belong
to the plane ${\cal M}_*$, will be denoted by the subscript $*$. For
example, the two Killings vectors are $\eta_0^i$, $\xi_0^i$. Note,
that for any pair $X_0^i, Y_*^i$ one has,
$$ X_0^i\,Y_*^k\,g_{ik} \equiv (X_0\,Y_*) = 0.$$

\subsection{Stationary and axially symmetric photosphere}

\subsubsection{The photosphere}

\noindent Numerical solutions of slim accretion disks provide the location of
the photosphere given by $H_{\rm Ph}(r)=r \cos\theta$. This may be put into
$r \cos\theta - H_{\rm Ph}(r) \equiv F(r, \theta) = 0$. The normal vector to the
photosphere surface has the following $[r,\theta]$ components,

\be
N_*^i=\tilde N_* \left[\pder F r,\pder F\theta\right]=\tilde N_*' \left[\der{\theta_*}{r},1\right],\ee
where
\be\der{\theta_*(r)}r=-\pder Fr\left /\pder F\theta\right.\ee
is the derivative of the angle defining the location of the photosphere at a given radial coordinate [$\cos\theta_*(r)=H_{\rm Ph}(r) / r$]. Its non-zero components after normalization [$(N_* N_*)=-1$] are

\be N_*^r=\der{\theta_*}r(-g_{\theta\theta})^{-1/2}\left[1+\frac{g_{rr}}{g_{\theta\theta}}\left(\der{\theta_*}r\right)^2\right]^{-1/2},\ee
\be\nonumber N_*^\theta=(-g_{\theta\theta})^{-1/2}\left[1+\frac{g_{rr}}{g_{\theta\theta}}\left(\der{\theta_*}r\right)^2\right]^{-1/2}.
\ee
There are two unique vectors $S_*$ confined in the $[r,\theta]$ plane which are orthogonal to $N_*$ (and therefore are tangent to the surface). From $(S_*N_*)=0$ and $(S_*S_*)=-1$ one obtains the non-zero components of one of them:

\be\label{e.Sstar}
 S_*^r=(g_{rr})^{-1}\left[-\frac1{g_{rr}}-\frac1{g_{\theta\theta}}\left(\der{\theta_*}r\right)^2\right]^{-1/2},\ee
\be\nonumber S_*^\theta=-(g_{\theta\theta})^{-1}\left(\der{\theta_*}r\right)\left[-\frac1{g_{rr}}-\frac1{g_{\theta\theta}}\left(\der{\theta_*}r\right)^2\right]^{-1/2}.\ee

\subsubsection{The four-velocity of matter and the tetrad}
\label{s.decomposition}
\noindent The four-velocity $u$ of gas  moving along the photosphere may be decomposed into

\be
u^i=\tilde A(u_0^i+vS_*^i),
\ee
where
\be
u_0^i=\tilde A_0\left(\eta^i+\Omega\xi^i\right)
\ee
is the four-velocity of an observer with azimuthal motion only. The normalization constant $\tilde A_0$ comes from $(u_0u_0)=1$ and equals
\be\tilde A_0=\left[g_{tt}+\Omega g_{\phi\phi}(\Omega-2\omega)\right]^{-1/2}.\ee
 It is useful to construct a spacelike vector ($\kappa_0$) confined in
 the $[t,\phi]$ plane, that is perpendicular both to $u$ and
 $u_0$. From $(\kappa \kappa)=-1$ and, e.g., $(\kappa u_0)=0$ we have,
\be\kappa_0^i=\frac{(l\eta^i+\xi^i)}{\left[-g_{\phi\phi}(1-\Omega l)(1-\omega l)\right]^{1/2}},\ee
where $l=u_\phi/u_t$ is the specific angular momentum. Note that the set of vectors $[u_0^i, ~N_*^i, ~\kappa_0^i, ~S_*^i]$ already forms the desired tetrad valid for the pure rotation ($u^r=0$) case.

\noindent The normalization condition $(uu)=1$ gives,

\be\tilde A=\left[g_{tt}+\Omega g_{\phi\phi}(\Omega-2\omega)-v^2\right]^{-1/2},\ee
where $v$ is related to the radial component of the gas four-velocity $u^r$ by:
\be v^2=\frac{\left(u^r/S^r_*\right)^2(g_{tt}+\Omega g_{\phi\phi} (\Omega - 2\omega))}{1+\left(u^r/S^r_*\right)^2}.\label{e.V}
\ee

\noindent The vectors we have just calculated ($u$, $\kappa_0$) are both orthogonal to $N_*$ since $(N_*S_*)=0$. To complete the tetrad we need one more spacelike vector ($S$) that is orthogonal to these three. Let us decompose it into,
\be S^i=\alpha u^i + \beta \kappa_0^i + \gamma N_*^i +\delta S_*^i. \ee
The orthogonality conditions $(\kappa_0 S)=0$ and $(N_* S)=0$ give immediately $\gamma=\beta=0$. The only non-trivial condition is $(u S)=0$. Together with $(SS)=-1$ it leads to:

\be S^i=(1+\tilde A^2v^2)^{-1/2}(\tilde A vu^i+S^i_*).\ee

\noindent The vectors $u^i,  ~N_*^i, ~\kappa_0^i,~S^i$ form an orthonormal tetrad in the Kerr spacetime:
\begin{equation} \label{velocity-tetrad}
e^i_{~(A)} = [u^i,~N_*^i,  ~\kappa_0^i, ~S^i].
\end{equation}
This tetrad {\it is known}
directly from the slim disk solutions, as it depends on the calculated quantities ($u^r$, $\Omega$, $l$ and $\theta_*(r)$) only.
Any spacetime vector $X^i$, could be uniquely
decomposed into this tetrad with,
$X_{(A)} = X_i\,e^i_{~(A)}$.

\subsection{The general case}

In this section we will assume nothing about the four-velocity of matter
$u^i$ and the location of photosphere. Both may be non-stationary and non-axially symmetric.
Following the same framework as in the previous sub-Section, we will describe how to obtain the tetrad of an observer
instantaneously located at the photosphere that depends only on the quantities calculated by accretion disk models.

\subsubsection{The four velocity}

\noindent Similarly as in Section \ref{s.decomposition}, we may always
{\it uniquely} decompose $u^i$, a general timelike unit vector, into
\begin{equation} \label{four-velocity2}
u^i = {\tilde A} (u_0^i + v S_*^i),
\end{equation}
\noindent where $u_0^i$ is a timelike unit vector, and $S_*^i$ is
a spacelike unit vector. Formula (\ref{four-velocity2}) uniquely
defines the two vectors $u_0^i, S_*^i$ and the two scalars
${\tilde A}, v$. The vectors and scalars
\begin{equation} \label{known-velocity}
\{ {\tilde A}, v, u_0^i, S_*^i \},
\end{equation}
can be calculated
from {\it known} quantities given by slim disk model solutions,

\noindent The four-velocity (\ref{four-velocity2}) defines also the
instantaneous 3-space of the comoving observer with the metric
$\gamma_{ik}$ and the projection tensor $h^{~i}_k$,
\begin{eqnarray}
\gamma_{ik} &=& g_{ik} - u_i\,u_k,\\
\label{comoving-metric}
h^i_{~k} &=& \delta^i_{~k} - u^i\,u_k
\end{eqnarray}
\noindent We define two unit vectors $\kappa_0^i$ and $N_*^i$ by
the unique condition,
\begin{equation} \label{kappa-N}
(\kappa_0 u_0) = 0, ~~(S_* N_*) = 0.
\end{equation}

\noindent As before, the four vectors
\begin{equation} \label{velocity-tetrad2}
e^i_{~(A)} = [u_0^i, ~\kappa_0^i, ~N_*^i, ~S_*^i],
\end{equation}
\noindent which form an orthonormal tetrad of an observer with the four-velocity $u_0^i$ can calculated from the solutions of the slim-disk equations.

\subsubsection{The photosphere}

\noindent In the most general case of a
non-stationary and non-axially symmetric photosphere, the location of the photosphere
may be described by the following condition,
\begin{equation} \label{general-photosphere}
F(t, \phi, r, \theta) = 0.
\end{equation}
\noindent The vector $\tilde N$ normal to the photosphere has
the components,
\begin{equation}
\label{normal-photosphere}
{\tilde N}^i = \left[
\frac{\partial F}{\partial t},
\frac{\partial F}{\partial \phi},
\frac{\partial F}{\partial r},
\frac{\partial F}{\partial \theta}
\right]
\end{equation}
\noindent which may be calculated from slim disk solutions.

\noindent Let us project ${\tilde N}$ into the instantaneous
3-space of the comoving observer (\ref{comoving-metric}) and
normalize to a unit vector after the projection,
\begin{equation}
\label{projected-normal}
N^i = \frac{{\hat N}^i}{\vert ({\hat N}{\hat N})\vert^{1/2}},
~~~{\hat N}^i = {\tilde N}^k\,h^i_{~k}.
\end{equation}
In terms of the tetrad (\ref{velocity-tetrad2}), such constructed
vector $N^i$ has the following decomposition,
\begin{equation}
\label{projected-normal-decomposition}
N^i = {\tilde N}\,[\alpha\, (u_0^i) + 1\, (N_*^i) + \gamma\,
(\kappa_0^i) + \delta\, (S_*^i)].
\end{equation}
\noindent The components ${\tilde N}, \alpha, \gamma, \delta$ are
known.

\subsubsection{The tetrad}

\noindent Let us now write decompositions of the four vectors: the
first two we have derived, the next two guessed (but the
guess should be obvious):
\begin{eqnarray}
u^i &=&{\tilde A}\,[1\, (u_0^i) \,+ 0\, (N_*^i) \,\,+ 0\,
(\kappa_0^i) \,+ V\, (S_*^i)], \label{final-tetrad-u}\\
N^i &=&{\tilde N}\,[\alpha\, (u_0^i) + 1\, (N_*^i) \,\,+ \gamma\,
(\kappa_0^i) \,+ \delta\, (S_*^i)], \label{final-tetrad-N}\\
\kappa^i &=&{\tilde \kappa}\,\,\,[0\, (u_0^i) \,+ b\, (N_*^i)
\,\,+
1\,(\kappa_0^i) \,+ 0\, (S_*^i)], \label{final-tetrad-K}\\
S^i &=&{\tilde S}\,\,[A\, (u_0^i) + B\, (N_*^i) + C\,(\kappa_0^i)
+ 1\, (S_*^i)] \label{final-tetrad-S}.
\end{eqnarray}
\noindent The four unknown components, $b, A, B, C$ one calculates
from the four non-trivial orthogonality conditions ($(u\kappa) \equiv 0$ by construction, cf. (\ref{final-tetrad-u})
and (\ref{final-tetrad-K})),
\begin{equation}
\label{orthogonality-conditions}
 (uS) = 0,~~ (NS) = 0,~~
(S\kappa) = 0,~~ (N\kappa) = 0,
\end{equation}
\noindent and the two unknown factors ${\tilde \kappa}, ~~ {\tilde
S}$, from the two normalization conditions
\begin{equation}
\label{normalization-conditions} (\kappa \kappa) = -1, ~~ (SS) =
-1.
\end{equation}
\noindent The conditions (\ref{orthogonality-conditions}),
(\ref{normalization-conditions}) are given by linear equations.

\noindent Equations
(\ref{final-tetrad-u})-(\ref{normalization-conditions}) define the
tetrad $e_{~(A)}^{~i}$ of an observer comoving with matter,
and instantaneously located at the photosphere:
\begin{equation}
\label{tetrad-final-final}
e_{~(A)}^{~i} = [u^i, ~N^i, ~\kappa^i, ~S^i].
\end{equation}
Both the matter
and the photosphere move in a general manner. The zenithal
direction in the local observer's sky is given by $N^i$.

\section{Integration over the world-tube of the
photosphere}
\label{ap.integration}

\noindent For stationary and axially symetric models we have so far,

\noindent $ N^i = N^i_* =~$ unit vector orthogonal to the photosphere. It
is in the $[r, \theta]$ plane

\vskip 0.1truecm

\noindent $ S_*^i =~$ unit vector orthogonal to $N^i$ that lives
in the $[r, \theta]$ plane

\vskip 0.1truecm

\noindent $ u^i =~$ four-velocity of matter. It lives in the $[t,
\phi, r, \theta]$ space-time

\vskip 0.1truecm

\noindent $ \kappa^i =\kappa_0^i=~$ unit vector orthogonal to $U^i$ that
lives in the $[t, \phi]$ plane

\vskip 0.1truecm

\noindent $ S^i =~$ unit vector orthogonal to $U^i$, $N^i$ and
$\kappa^i$. It lives in the $[t, \phi, r, \theta]$ space-time

\vskip 0.1truecm

\noindent $ e_{~(A)}^{~i} =~$ $[ u^i, N^i,\kappa^i,  S^i] =~$ the
tetrad comoving with an observer located in the photosphere

\vskip 0.2truecm

\noindent The integration of a vector $(...)_i$ over the 3-D
hypersurface ${\cal H}$ orthogonal to $N^i$ (i.e. the 3-D
world-tube of the photosphere) may be symbolically written as

\be \int_{\cal H} (...)_i N^i dS, \ee  \vskip 0.1truecm

\noindent where $dS$ is the ``volume element'' in ${\cal H}$.

\vskip 0.2truecm

\noindent Obviously, the hypersurface ${\cal H}$ is spanned by the
three vectors $[u^i, \kappa^i, S^i]_N$. Each of them is a linear
combination of $[\eta^i, \xi^i, S_*^i]_{N}$, and each of
the three vectors from $[\eta^i, \xi^i, S_*^i]_{N}$ is
orthogonal to $N^i$.

\vskip 0.2truecm

\noindent Therefore, one may say that the hypersurface ${\cal H}$
is spanned by $[\eta^i, \xi^i, S_*^i]_{N}$. It will be
convenient to write

\be dS = dA dR, ~~~dR = dr\sqrt{ g_{rr} +
g_{\theta \theta}\left(\frac{d\theta_*}{dr}\right)^2}, \ee

\noindent where $dR$ is the line element along the vector $S_*^i$,
i.e. along the photosphere in the $[r, \theta]$ plane, with
$\theta = \theta_*(r)$ defining the location of the photosphere,
and $dA$ is the surface element on the $[t, \phi]$ plane.

\vskip 0.2truecm

\noindent In order to calculate $dA$, imagine an infinitesimal
parallelogram with sides that are located along the $t =$~const
and $\phi =$~const lines. The proper lengths of the sides are  $du
= \vert g_{tt}\vert^{1/2}dt$ and $dv = \vert g_{\phi
\phi}\vert^{1/2}d\phi$ respectively, and therefore $dA$, which is
just the area of the parallelogram, is given by

\be dA = du\, dv \sin \alpha = dt\,d\phi \,\vert g_{tt}\vert^{1/2}\vert g_{\phi
\phi}\vert^{1/2} \sin \alpha  \ee

\noindent where $\alpha$ is the angle between the two sides.
Obviously, the cosine of this angle is given by the scalar product
of the two unit vectors $n_i$ and $x_i$ pointing in the $[t,
\phi]$ plane into $t$ and $\phi$ directions respectively. These
vectors are given by (note that $n_i =~$ ZAMO),

\be n_i = \frac{(\nabla_i t)}{\vert g^{jk} (\nabla_j t)( \nabla_k
t)\vert^{1/2}}, ~~~x_i = \frac{(\nabla_i \phi)}{\vert g^{jk}
(\nabla_j \phi)( \nabla_k \phi) \vert^{1/2}}.\ee

\noindent Because $(\nabla_i t) = \delta^t_{~i}$ and $(\nabla_i
\phi) = \delta^{\phi}_{~i}$, one may write,

\be n_i = \frac{\delta^t_{~i}}{\vert g^{tt}\vert^{1/2}}
, ~~~x_i = \frac{\delta^\phi_{~i}}{\vert g^{\phi
\phi}\vert^{1/2}}.\ee
Therefore,
\be  \cos \alpha = n_i x_k g^{ik} = \frac
{g^{t\phi}}{\vert g^{tt}\vert^{1/2}\vert g^{\phi \phi}\vert^{1/2}}
= - \frac {g_{t\phi}}{\vert g_{tt}\vert^{1/2}\vert g_{\phi
\phi}\vert^{1/2}},\ee
and
\be \sin \alpha = \frac{(g_{t\phi}^2 - g_{tt}\,g_{\phi\phi})^{1/2}}{\vert g_{tt}\vert^{1/2}\vert g_{\phi
\phi}\vert^{1/2}}\ee

\noindent Inserting this into the formula for $dA$ we get $dA =
dt\,d\phi\,(g_{t\phi}^2 - g_{tt}\,g_{\phi\phi})^{1/2}$. The final formula for $dS$ is,

\be
\label{ap.dS}
{\rm d}S = {\rm d}t\,{\rm d}\phi\,{\rm d}r\,\left( g_{t\phi}^2 -
g_{tt}\,g_{\phi\phi}\right )^{1/2}\,\sqrt{ g_{rr} + g_{\theta
\theta}\left(\frac{d\theta_*}{dr}\right)^2}.\ee

\end{appendix}

\end{document}